\documentclass[twocolumn,prb,aps,10pt]{revtex4-1}

\usepackage{amsfonts,amssymb}
\usepackage[sumlimits,intlimits]{amsmath}
\usepackage{graphics}
\usepackage{graphicx}
\usepackage{epsfig}
\usepackage{mathrsfs}
\usepackage{wasysym}
\usepackage{textcomp}
\usepackage{verbatim}
\usepackage{hyperref}    
\usepackage{bm}          

\newcommand{\be}{\begin{equation}}
\newcommand{\ee}{\end{equation}}
\newcommand{\Tes}{T_\text{ES}}
\newcommand{\e}{\varepsilon}

\begin{document}

\title{Anomalously small resistivity and thermopower of strongly compensated semiconductors and topological insulators}

\author{Tianran Chen}
\author{B. I. Shklovskii}
\affiliation{Fine Theoretical Physics Institute, University of Minnesota, Minneapolis, MN 55455, USA}

\date{\today}

\begin{abstract}
In the recent paper, we explained why the maximum bulk resistivity of topological insulators (TIs) such as Bi$_{2}$Se$_{3}$ is so small~\cite{Skinner2012wit}. Using the model of completely compensated semiconductor we showed that when the Fermi level is pinned in the middle of the gap the activation energy of resistivity $\Delta =0.3 (E_g/2)$, where $E_g$ is the semiconductor gap. In this paper, we consider a strongly compensated $n$-type semiconductor. We find the position of the Fermi level $\mu$ calculated from the bottom of the conduction band $E_c$ and the activation energy of resistivity $\Delta$ as a function of compensation $K$, and show that $\Delta = 0.3 (E_c-\mu) $ holds at any $0< 1-K \ll 1$. In the same range of relatively high temperatures, the Peltier energy (heat) $\Pi$ is even smaller: $\Pi \simeq \Delta/2 = 0.15(E_c - \mu)$. We also show that at low temperatures, the activated conductivity crosses over to variable range hopping (VRH) and find the characteristic temperature of VRH, $\Tes$, as a function of $1-K$. 
\end{abstract}
\maketitle

\section{Introduction}

The three-dimensional (3D) topological insulator (TI)~\cite{Fu2007tii, Moore2007tio, Fu2007tiw, Qi2008tft, Roy2009tpa} has gapless surface states, which host a spectrum of quantum transport phenomena~\cite{Hasan2010c:t, Qi2011tia}. While a number of crystals have been identified to be 3D TIs, most of them are poor insulators and the bulk of TI crystals of substantial size ($> 10 \ \mu$m) shunts the surface conductivity. The current literature~\cite{Qu2010qoa, Analytis2010tss, Checkelsky2009qii, Butch2010sss, Analytis2010bfs, Eto2010aoo, Ren2011oot, Ren2011o$s, Ren2012flt}broadly discusses how one can achieve a bulk-insulating state. 

Typically as-grown TI crystals such as Bi$_{2}$Se$_{3}$ are heavily doped $n$-type semiconductors. (It is believed that Bi$_{2}$Se$_{3}$ is doped by Se vacancies.) To make them insulating, these TIs are compensated by acceptors. With increasing compensation $K =N_A/N_D$, where $N_D$ and $N_A$ are the concentrations of monovalent donors and acceptors, the Fermi level shifts from the conduction band to inside the gap and then into the valence band at $K > 1$. When compensation of donors is complete, $K=1$, the Fermi level is in the middle of the gap and the most insulating state of TI is achieved. For a TI with a gap $E_g \sim 0.3$ eV the resistivity is expected to obey the activation law 
\be
\rho = \rho_0 \exp(\Delta/k_B T)
\label{act}
\ee
with activation energy $\Delta = E_g/2 \sim 0.15$ eV, so that the TI is a good insulator at room temperatures and below. 

However, the current experimental situation near $K=1$ is frustrating~\cite{Ren2011o$s}. In the temperature range from 100 and 300 K, although resistivity is activated, the activation energy $\Delta \sim 50$ meV, which is three times smaller than expected. At $T \sim 100$ K the activated transport crosses over to variable range hopping (VRH). When temperature is further decreased, resistivity grows even more slowly and below 50 K, resistivity saturates around $\rho(T) < 10 \ \Omega$cm. This means that in spite of complete compensation, even at helium temperatures conductance of TI samples thicker than $10 \ \mu$m is dominated by the bulk. 

\begin{figure}[htb!]
\centering
\includegraphics[width=0.44 \textwidth]{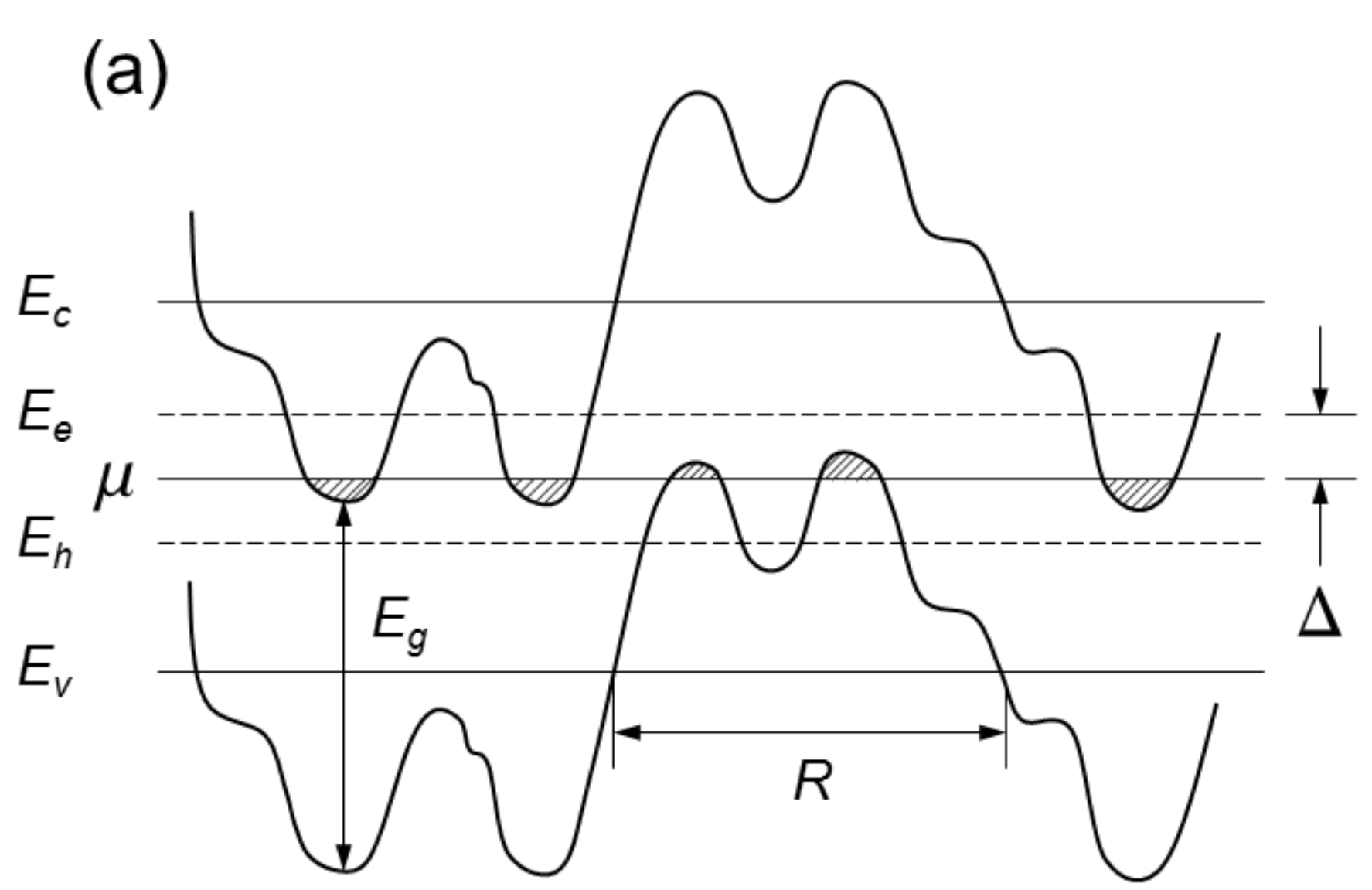}

\includegraphics[width=0.44 \textwidth]{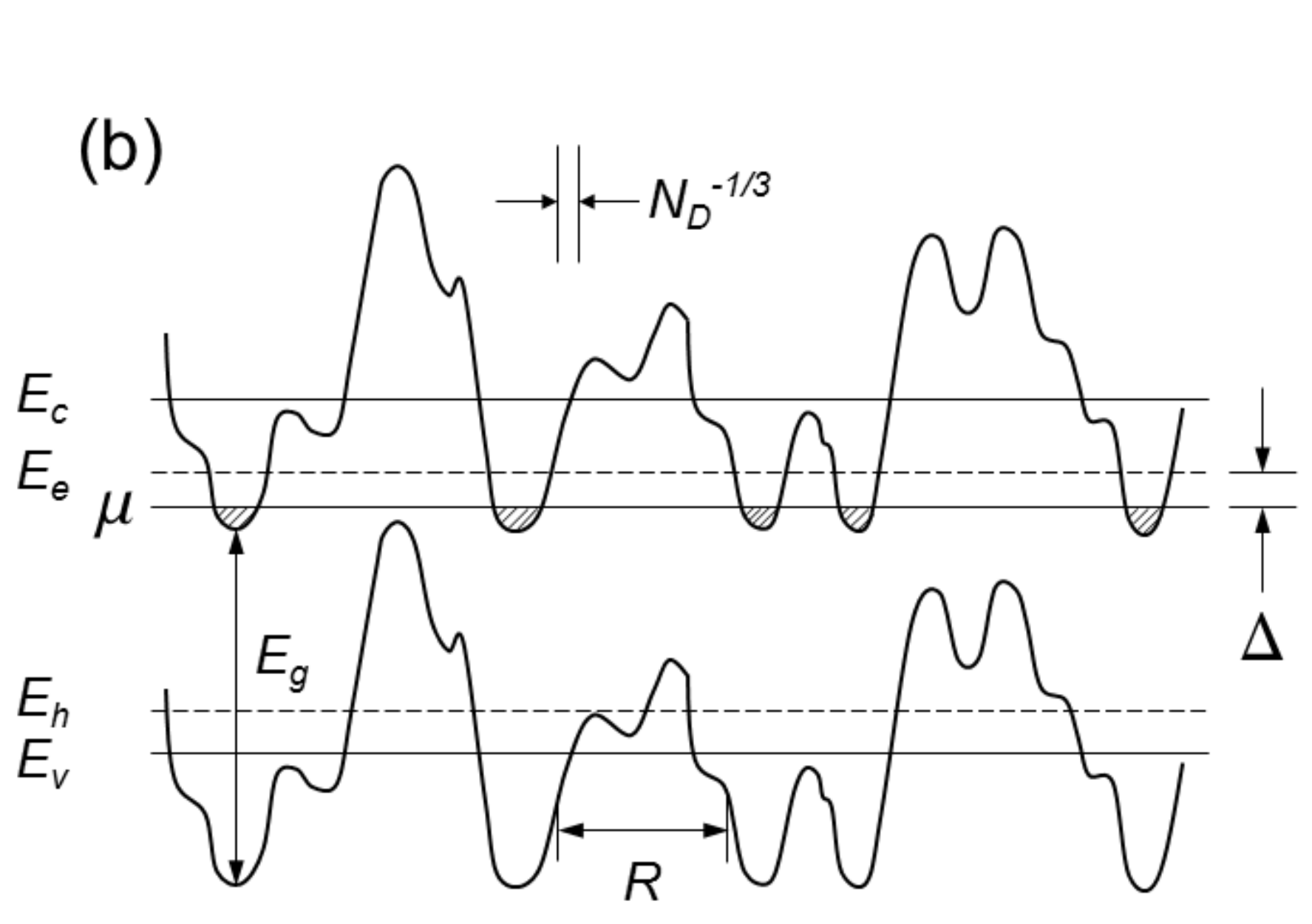}
\caption{Energy diagram of (a) completely compensated semiconductor ($K = 1$) and (b) strongly compensated semiconductor ($1-K \ll 1$) with gap $E_g$. The upper and  the lower straight lines indicate the unperturbed positions of bottom of the conduction band, $E_c$, and ceiling of the valence band $E_{\nu}$; the middle straight line corresponds to the Fermi level $\mu$. Meandering lines represent the band edges, which are modulated by the fluctuating potential of charged impurities. $R$ is the characteristic size of potential fluctuations. Percolation levels $E_e$ for electrons and $E_h$ for holes are shown by dashed lines. Puddles occupied by carriers are shaded. Shallow impurities levels are not shown because they practically merge with band edges.}
\label{fig:band}
\end{figure}

In the recent paper~\cite{Skinner2012wit}, we suggest an explanation of anomalously large bulk conductivity of TI at $K=1$. We assume that both donors and acceptors are shallow and randomly positioned in space 
and we use the theory of completely compensated semiconductor (CCS)~\cite{Efros1984epo, Shklovskii1972ccc}. 
The idea that at $K=1$, when almost all donors and acceptors are charged, random spatial fluctuations in the local concentration of impurities lead to large fluctuations of charge. Because the average concentration of screening electrons $n = N_D-N_A \ll N_D$, the random potential is poorly screened and has huge fluctuations. These fluctuations bend conduction and valence band edges and in some places bring them to the Fermi level, creating electron and hole puddles, which in turn non-linearly screen the random potential. Therefore, the amplitude of potential fluctuations is limited by $E_g/2$. The ground state of the completely compensated semiconductor shown in Fig. \ref{fig:band}a therefore reminds a network of $p$-$n$ junctions~\cite{Shklovskii1972ccc,Efros1984epo}. The characteristic size $R$ of these $p$-$n$ junctions in Bi$_{2}$Se$_{3}$ with $E_g \simeq 0.3 \ eV$, $N_D = 10^{19}$ cm$^{-3}$, and dielectric constant $\kappa = 30$ is $R \simeq 150$ nm $\gg N_D^{-1/3} = 4.6$ nm~\cite{Skinner2012wit}, i.e., we deal with a very long range potential. As a result, the resistivity can be dramatically different from the one for the flat bands picture of TI~\cite{Skinner2012wit}. First, at relatively high temperatures, the activated conduction is due to the electrons and holes being activated from the Fermi level to their corresponding classical percolation levels (classical mobility edges), $E_e$ and $E_h$, in the conduction and the valence bands. According to numerical modeling ~\cite{Skinner2012wit} at $K = 1$ the activation energy $\Delta \simeq 0.15 E_g$, because $E_e$ and $E_h$ are substantially closer to the Fermi level $\mu$ than the unperturbed by a random potential bottom of the conduction band $E_c$ and ceiling of the valence band $E_{\nu}$ [see Fig. \ref{fig:band}a]. Second, at low enough temperatures, electrons and holes can hop (tunnel) between puddles, so that variable range hopping replaces activated transport. We showed that the activated resistivity crosses over directly to Efros-Shklovskii (ES) law of VRH ~\cite{Efros1975cga},
\be
\rho = \rho_{0}\exp(\Tes/T)^{1/2},
\label{eslaw}
\ee
where $\Tes = Ce^2/k_B\kappa\xi$, $e$ is the electron charge, $\xi$ is the localization length of the states with the Fermi level energy and $C=4.4$ is a numerical coefficient. Together, our results for the activated and VRH resistivity established the universal upper limit of resistivity $\rho(T)$ one can achieve for a 3D TI compensated by shallow inpurities.

In this paper, we change our focus from a possible maximum bulk resistivity of a completely compensated semiconductor at $K=1$ to the more practical question of the dependence of bulk resistivity of a strongly compensated semiconductor (SCS) on $K$ at $0 < 1-K \ll 1$. Indeed, with existing methods of growth of TI samples one can not get $K=1$ exactly. It is important to know how stable the resistivity results at $K=1$ are for the case of $1-K \ll 1$. For example, one can ask at which $1-K$ the activation energy $\Delta$ is twice smaller than at $K=1$. For definiteness, we consider $n$-type SCS, where the concentration of electrons $n = N_D - N_A \ll N_D$ and  $1-K \ll 1$. We model numerically the ground state of such SCS and its resistivity using algorithms similar to Ref.~\cite{Skinner2012wit}. We find that in agreement with the analytic theory~\cite{Efros1984epo}, when $1-K$ grows, the screening of the random potential improves and its correlation length $R$ decreases. The amplitude of the random potential decreases as well. As a result, hole puddles shrink and eventually vanish and the chemical potential $\mu$ moves up, so that $E_c-\mu$ decreases. One can say that with increasing  $1-K$, the screening due to bending of the conduction band occurs only while all acceptors remain occupied by electrons and negatively charged. All these changes are illustrated by transition from (a) to (b) in Fig. \ref{fig:band}. 

As a result of these changes, the activation energy $\Delta$ decreases with growing $1-K$. We find that the relation $\Delta = 0.3 (E_c-\mu) $ obtained in Ref.~\cite{Skinner2012wit} for $K=1$  remains valid for $1-K \ll 1$ (see Fig. \ref{fig:EA} below) as well. [In $p$-type semiconductor where $K=N_D/N_A$, a similar relationship $\Delta = 0.3 (\mu-E_v)$ takes place.] By $K=0.97$ the activation energy $\Delta$ is about two times smaller than at $K=1$. This result shows that achieving maximum resistivity with $\Delta = 0.15 E_g$ is problematic. It also explains the origin of large scatter of magnitude of $\Delta$ among TI samples~\cite{Ren2011o$s}.

In principle, our prediction that $\Delta = 0.3 (E_c - \mu)$ can be directly compared with experiments in TIs. Indeed, for each $K$, the position of the Fermi level can be found via measurements of the surface concentration of electrons in the gapless surface state using Shubnikov-de-Haas oscillations. On the other hand, at low temperatures, we find numerically a direct cross-over from activation to ES VRH. We also find how $\Tes$ being correlated with $\Delta$ decreases with $1-K$. 

Our assumption of random distribution of impurities is crucial for this theory. Usually, for samples made by cooling from melt, the distribution of impurities in space is a snapshot of the distribution the impurities have at higher temperature when the diffusion of impurities practically freezes. In semiconductors with a narrow enough gap at this temperature, there is a concentration of intrinsic carriers larger than the concentration of impurities. Intrinsic carriers screen the Coulomb interaction between impurities, so that impurities remain randomly distributed in space. At lower temperatures, when intrinsic carriers recombine, impurities are left in random positions~\cite{Galperin1972Sov, Efros1984epo}. 
If diffusion freezes at $T \sim 1000K$, it is reasonable to assume that impurities are randomly positioned in a semiconductor with $E_g \leq 0.3 \ eV$. This justifies the use of this theory for typical TIs. Our results are applicable to other narrow gap semiconductors, for example, InSb. (Historically, large effort was made to make InSb insulating via strong compensation. The goal was to improve characteristics of InSb based photo-detectors. Results were again frustrating: the dark resistivity was too small. Our results are in reasonable agreement with transport experiment data for InSb~\cite{Gershenzon1975Sov, Yaremenko1975}.)

The plan of the paper is as follows. In Sec.\ II, we formulate the model, explain the algorithm of numerical simulation of the pseudoground state and present results for the density of states (DOS). In Sec.\ III, we present our algorithm for computation of hopping conductivity, analyze our results and arrive at a small activation energy for conduction band resistivity $\Delta = 0.3 (E_c-\mu)$. We also evaluate the localization length of states with energy close to Fermi energy and estimate the characteristic temperature of ES law $\Tes$. In Sec.\ IV, we estimate the thermopower of strongly compensated semiconductor and show that the Peltier energy (heat) is $\Pi \simeq \Delta/2 = 0.15(E_c - \mu)$, in qualitative agreement with a recent experimental paper~\cite{Akrap2012tp}. We conclude in Sec.\ V, where we comment on predictions of this model for the Hall effect measurements and compare these predictions with experimental data~\cite{Ren2011o$s}.

\section{The model, pseudoground states, and the density of states}

To model a heavily doped SCS, we create a cube filled with 20000 donors and $20000K$ acceptors that are randomly positioned in space. We numerate all donors and acceptors by index $i$ and use $n_i = 0$ or $1$ for the number of electrons residing on a donor or an acceptor. In addition, we use a variable $f_i$ to discriminate between donors ($f_i = 1$) and acceptors ($f_i = -1$). The Hamiltonian of our system is 
\be
H= \sum_i \frac{E_g}{2} f_i n_i +  \sum_{\langle ij\rangle} V(r_{ij}) q_i q_j,
\label{Hamiltonian}
\ee
where $q_i = (f_i + 1)/2 - n_i$ is the net charge of site $i$ and all energies are defined relative to the Fermi level. The first term contains the energies of shallow donors and acceptors, which is very close to the semiconductor gap $E_g$. The second term of $H$ is the sum of interaction energies of charged impurities. If two impurities are at distance $r >> a_B$, where $a_B$ is the Bohr radius of impurity states, one can use the Coulomb interaction $V(r)=e^2/\kappa r$. For a pair of empty donors, one donor shifts down the energy of the electron on the other by an energy $V(r)= - e^2/\kappa r$. This classical form for $V(r)$ is good for a lightly doped SCS. But in a heavily doped SCS, where $a_B  > N_{D}^{-1/3}$, most impurities have at least one neighbor at distance $r < a_B$ and quantum-mechanical averaging over electron wave function becomes important. (This is why an uncompensated heavily doped semiconductor is a good metal). For example, such a pair of donors cannot create a state deeper than that of the helium-like ion with a binding energy $4E_B$, where $E_B = e^2 /2\kappa a_B$ is the binding energy of the shallow donor state. Here, we deal with heavily doped SCS, where $(E_c-\mu) > 4E_B$ and quantum effects limit the role of short-range potential. To model such a case, we continue to use the classical Hamiltonian Eq.\ (\ref{Hamiltonian}), but truncate the Coulomb potential to $V(r)=e^2/\kappa (r^2 + a^2_B)^{1/2}$. Note that Eq.\ (\ref{Hamiltonian}) does not include the kinetic energy of electrons and holes in conduction and valence bands and, therefore, aims only at description of the low temperature $(k_B T \ll E_g)$ physics of SCS. 

Below, we use dimensionless units for $r$, $a_B$, $H$, $E_g$, and $k_B T$, measuring all distances in units of $N_{D}^{-1/3}$ and all energies in units of $e^2 N_{D}^{1/3}/\kappa$. Thus Eq.\ (\ref{Hamiltonian}) now can be understood as dimensionless, where  $E_g \gg 1$ and $V(r)= (r^2 + a_B^2)^{-1/2}$. For TI with $E_g = 0.3$ eV, $\kappa = 30$, and $N_{D}=10^{19}$ cm$^{-3}$, we have $N_{D}^{-1/3} = 4.6$ nm and $e^2 N_{D}^{1/3}/\kappa \simeq 10$ meV, so that the dimensionless gap $E_g = 30$. We could not model $E_g = 30$, because in this case, the very large correlation length of long-range potential, $R$, leads to large size effect. Instead, we run more modest $E_g = 15$, for which the size effect requires extrapolation only at $K=1$~\cite{Skinner2012wit}. Our goal is to find the activation energy $\Delta$ and estmate $\Tes$ as a function of $K$ or $\mu$.
  
We search for the set $\{n_i,f_i\}$ that minimizes $H$ and use such a set to calculate the DOS and the conductivity. We start from the neutral system of all populated by electrons (negatively charged) acceptors ($n_i=1, q_i=-1$), of equal number of randomly chosen $20000K$ empty (positively charged) donors ($n_i=0, q_i=1$), and of $20000(1-K)$ filled (neutral) donors ($n_i= 1, q_i= 0$). Charged donors and acceptors create a random potential whose magnitude exceeds $E_g$. In order to screen the Coulomb potential fluctuations, some electrons leave acceptors for donors. At any stage of this process, there are two types of occupied states -- neutral donors and negatively charged acceptors, and two types of empty states -- positively charged donors and neutral acceptors, respectively. Electrons may hop from an occupied impurity to an empty one. If the proposed move lowers the total system energy $H$, then it is accepted, otherwise it is rejected. To check whether $H$ goes down, for a given set of electron occupation numbers $\{n_i, f_i\}$, it is convenient to introduce the single-electron energy state, $\e_i$, at a given impurity $i$:
\be
\e_i = \frac{E_g}{2}f_i -  \sum_{j \neq i} V(r_{ij})q_j.  
\label{se}
\ee
For all $i$, $j$ with $n_i = 1$ and $n_j = 0$, we check that ES pseudoground state stability criterion is satisfied:
\be
\e_j - \e_i - V(r_{ij}) > 0.
\label{eq:ES} 
\ee
If this criterion is not satisfied, we move the electron from impurity $i$ to $j$ and recalculate all $\e_i$. This process is done by looping all possible pairs of impurities $i, j$ with $n_{i} = 1$ and $n_{j} = 0$ and is continued until no single-electron transfers can be made to lower $H$. The final arrangement of electrons can be called a pseudoground state, because the higher stability criteria of ground state are not checked. Pseudoground states are known to describe the properties of the real ground state pretty well~\cite{Efros1984epo, Mobius1992cgi}. The results below are obtained at $E_g=15$, $a_B =1$ for $K=1, 0.99, 0.98, 0.97, 0.96$, and $0.95$ (averaged over 100 realizations of impurities coordinates).

\begin{figure}[tb!]
\centering
\includegraphics[width=0.48 \textwidth]{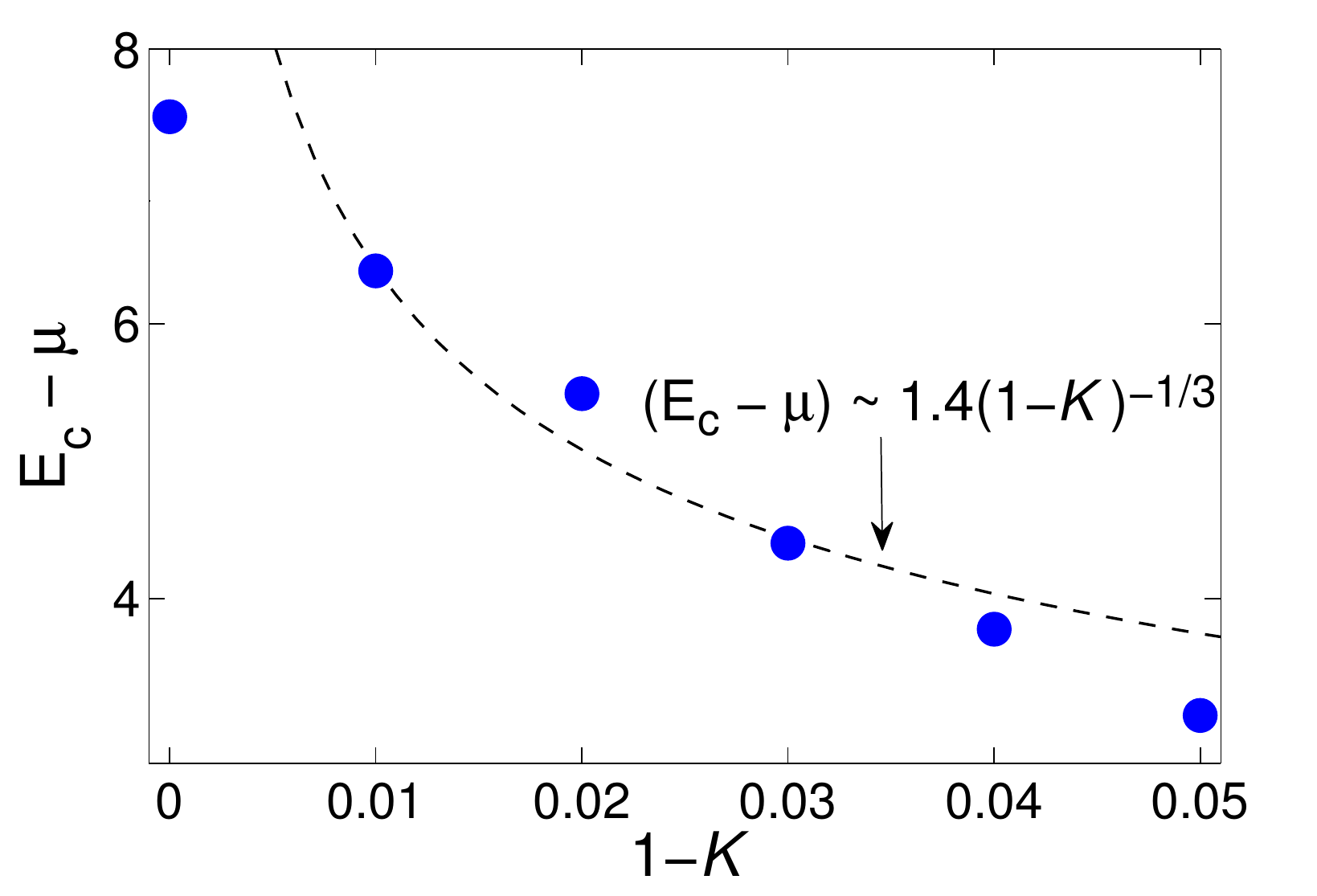}
\caption{(Color online) Fermi level $\mu$ as a function of $1-K$ for $a_B = 1$ and $E_g = 15$. The size of dots characterizes the uncertainty.}
\label{fig:MU}
\end{figure}

For a pseudoground state, we find the Fermi energy $\mu$ as a half distance between the minimum empty and maximum occupied energy $\e$. 
Fig. \ref{fig:MU} shows how the Fermi level $\mu(K)$ shifts from the middle of the gap towards the conduction band bottom with growing $1-K$.
At $1-K > 0.01$, this dependence is in reasonable agreement with the prediction of single-band theory (the theory that ignores valence band and acceptors) ~\cite{Efros1984epo} that $E_c-\mu = A (1-K)^{-1/3}$. However, note that for heavily doped SCS, the coefficient $A_h \simeq 1.4$ is twice smaller than the coefficient $A_l \simeq 2.8$ obtained  in Ref.~\cite{Efros1984epo} for a lightly doped SCS, where $a_B \ll 1$. In this case, the short-range Coulomb interaction at distance $r \ll N_D^{-1/3}$ leads to an additional contribution to $\mu$ of the same order of magnitude.

To confirm our understanding of results for $1-K > 0.01$, we obtained the same results for the position of Fermi level $\mu$ (and DOS of donors and conductivity, see below) using a simplified one-band model
where all acceptors are assumed to be negative.
Such program is similar to the classical impurity band program used in
Chapter 14 of Ref.~\cite{Efros1984epo}, but uses the redefined $V(r)$.

\begin{figure}[tb!]
\centering
\includegraphics[width=0.48 \textwidth]{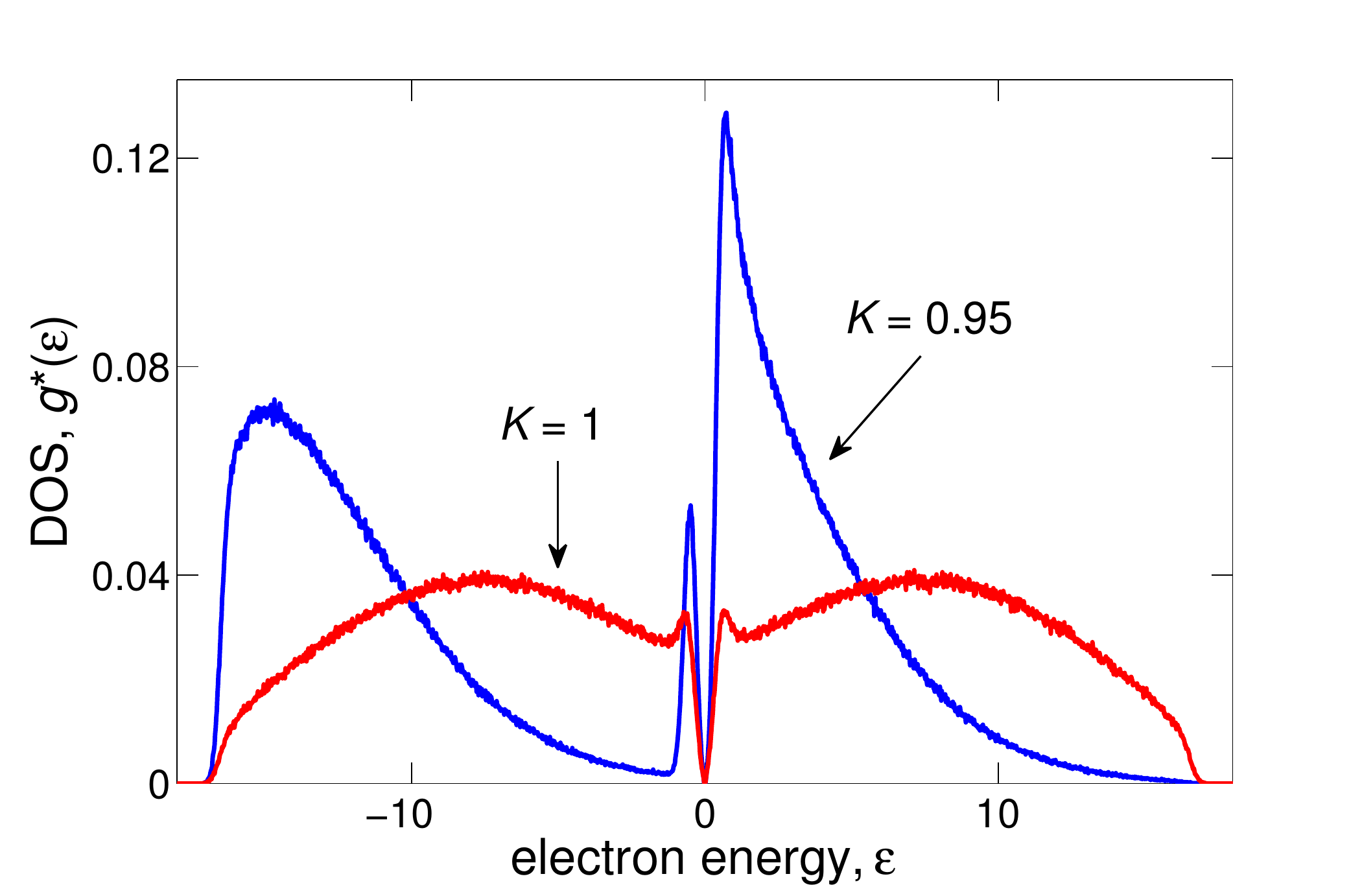}
\caption{(Color online) Dimensionless single-electron DOS $g^*(\e)$ in units of $[(1+K)N_D/(e^2N_D^{1/3}/\kappa)]$ as a function of $\e$ calculated from the Fermi level for $a_B = 1$ and $E_g = 15$ at $K$ = 0.95 (blue) and 1 (red). Impurity states with $\e <0$ are occupied and with  $\e  > 0$ are empty.
At $K = 1$, the total DOS of impurities has donor-acceptor symmetry, which is lost with growing $1-K$.}
\label{fig:DOS}
\end{figure}

The resulting DOS of impurities is shown in Fig. \ref{fig:DOS} for $K = 1$ and $K=0.95$. At $K = 1$, the almost constant symmetric DOS between $-E_g = -15$ and $E_g = 15$ reflects a practically uniform distribution of random potential from $-E_g/2$ to $E_g/2$, and a corresponding uniform distribution of band edges $E_c$ between $0$ and $E_g$ and $E_{\nu}$ between $0$ and $-E_g$ [see Fig. \ref{fig:band}(a)]. In the middle (at the Fermi level) one sees the ES Coulomb gap~\cite{Efros1975cga}. 

At $K < 1$, the DOS of impurities loses the donor-acceptor symmetry it has at $K=1$. As mentioned in Introduction (see Fig. \ref{fig:band}), with growing $1-K$, acceptors become all filled and disengaged from screening. Acceptor DOS  (leftmost peak) splits from the donor one, which in turn has two peaks separated by the Fermi level. The large right peak belongs to empty donors, while the small and narrow left peak belongs to occupied donors. The donor peaks are separated by the ES Coulomb gap. 

\begin{figure}[tb!]
\centering
\includegraphics[width=0.48 \textwidth]{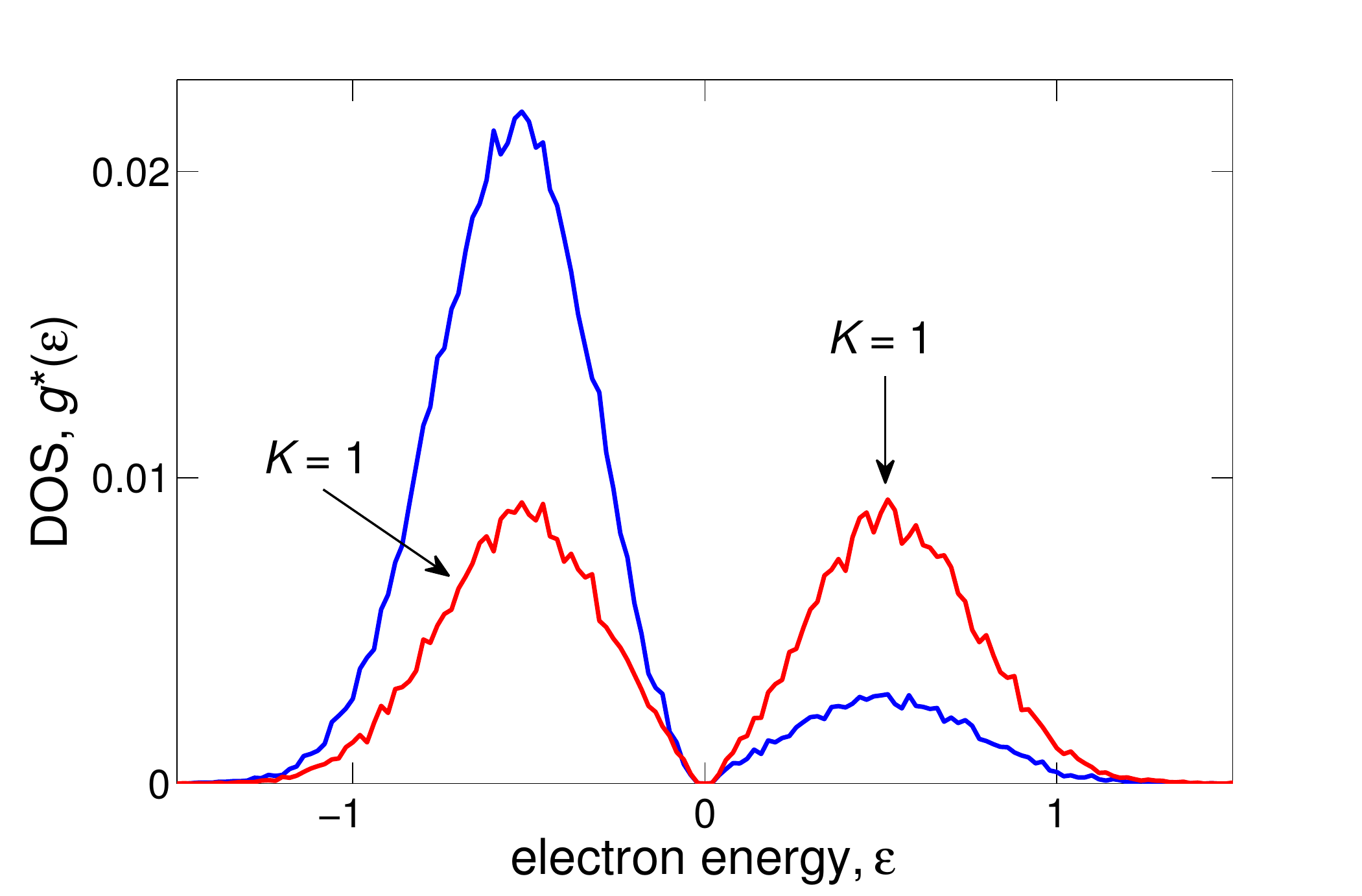}
\caption{(Color online) Dimensionless DOS $g^*(\e)$ for neutral (occupied by electrons) donors with $\e <0$ and neutral (empty) acceptors with $\e >0$ for $a_B = 1$ and $E_g = 15$ at $K$ = 0.98 (blue) and 1 (red).}
\label{fig:DOSEH}
\end{figure}

Growing with $1-K$ the disengagement of acceptors from screening is also illustrated in  Fig. \ref{fig:DOSEH}, where we show the DOS $g^*(\e)$ for neutral donors and acceptors. If at $K=1$, the total number of electrons and holes in puddles are equal, with growing $1-K$, the total number of electrons in electron paddles grows, while the total number of holes in hole puddles decreases. Thus, at $1-K \geq 0.02$, valence band practically plays no role in screening.

\section{Numerical modeling of hopping conductivity}

Once the energies $\{\e_i\}$ are known, we evaluate the resistivity using the approach of the Miller-Abrahams resistor network~\cite{Efros1984epo}. Each pair of impurities $i, j$ is said to be connected by the resistance $R_{ij}=R_0\exp[2 r_{ij}/\xi + \e_{ij}/k_BT]$, where the activation energy $\e_{ij}$ is defined~\cite{Efros1984epo} as follows: 
\be 
\e_{ij} = \left\{
\begin{array}{lr}
|\e_j - \e_i| - V(r_{ij}), &  \e_j\e_i < 0, \vspace{2mm} \\
\max \left[ \left|\e_i \right|, \left|\e_j \right| \right], &  \e_j\e_i > 0.
\end{array}
\right.
\label{eq:Eij}
\ee
The resistivity of the system as a whole is found using a percolation approach.  Specifically, we find the minimum value $R_c$ such that if all resistances $R_{ij}$ with $R_{ij} < R_c$ are left intact, while others are eliminated (replaced with $R = \infty$), then there exists a percolation pathway connecting opposite faces of the simulation cube. The system resistivity $\rho(T)$ is defined as $R_{c} N_D^{-1/3}$. Here, we concentrate on the exponential term of resistivity $\rho$ ignoring details of the prefactor~\cite{Efros1984epo}.

\begin{figure}[tb!]
\centering
\includegraphics[width=0.5 \textwidth]{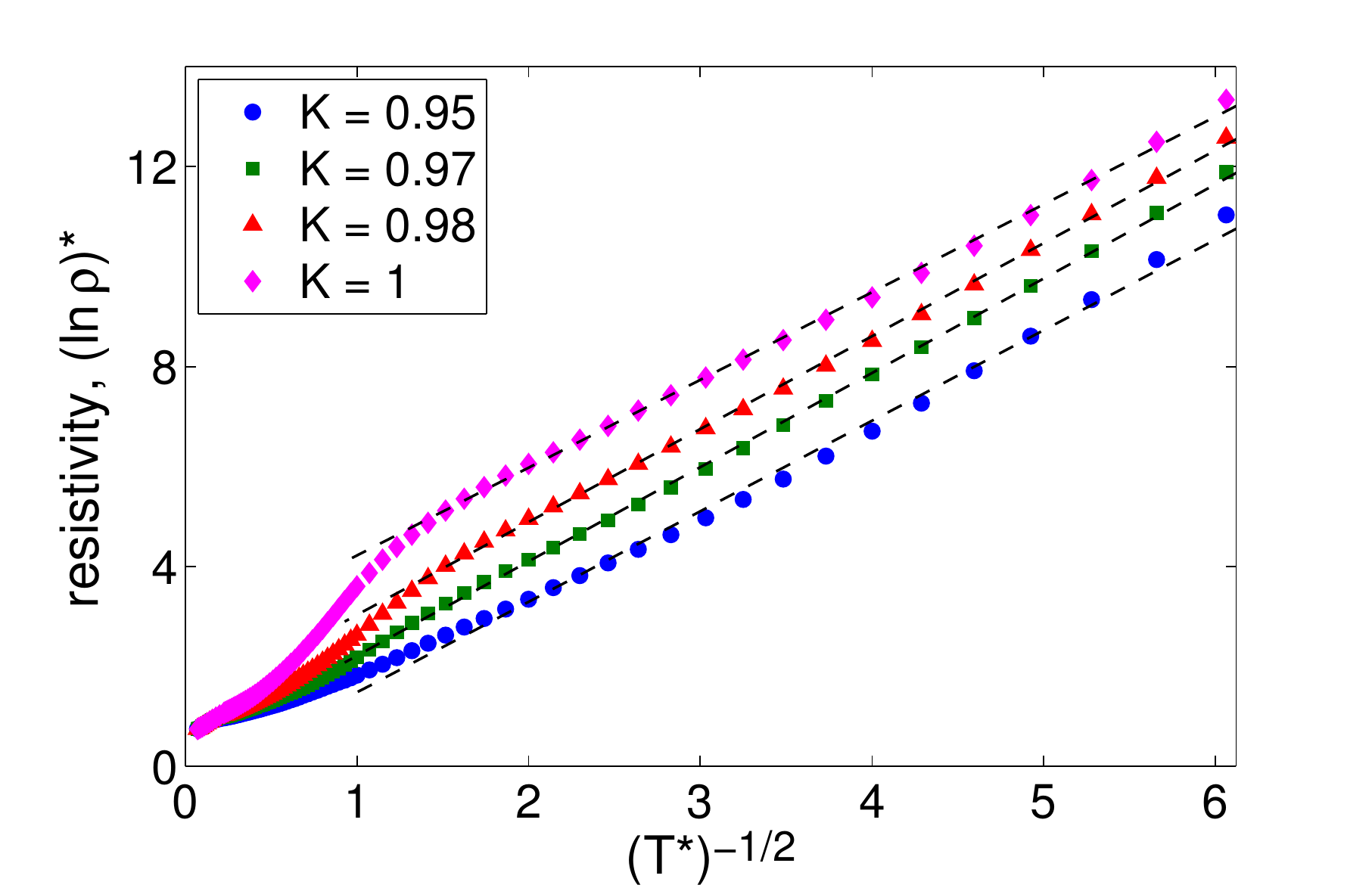}
\caption{(Color online) The temperature dependence of the resistivity in the whole temperature range $0.03 < T^* < 200$. The dimensionless resistance $(\ln \rho)^*$ is plotted against $(T^*)^{-1/2}$ to illustrate that the resistivity follows the ES law at low temperatures. The dashed lines are the best linear fits.}
\label{fig:RL}
\end{figure}

\begin{figure}[tb!]
\centering
\includegraphics[width=0.5 \textwidth]{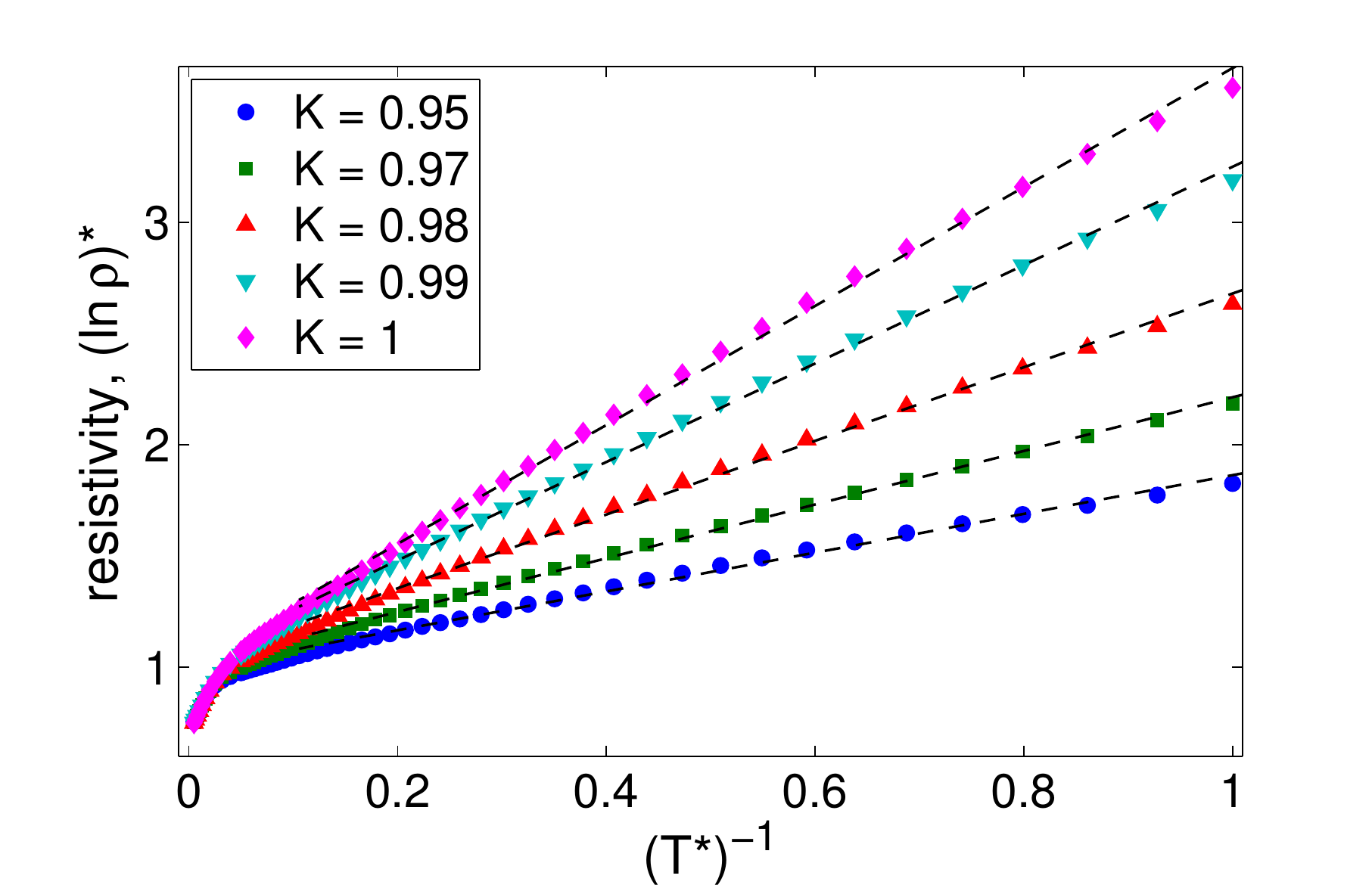}
\caption{(Color online) The temperature dependence of the resistivity in the high temperature range $1 < T^* < 200$. $(\ln \rho)^*$ is plotted against $(T^*)^{-1}$ to illustrate that the resistivity is activated at high temperatures. The dashed lines are the best linear fits.}
\label{fig:RH}
\end{figure}

For $K$ = 0.95, 0.97, 0.98, and 1 at $a_B = 1$ and $E_g=15$, the computed dependence of $(\ln\rho)^* = (\xi/2) \ln(R_c/R_0)$ is shown as a function of $(T^*)^{-1/2}$ in the huge range of temperatures $0.03 < T^* < 200$ in Fig. \ref{fig:RL}. Here, $T^* = 2 k_B T / \xi$ is yet another dimensionless temperature. These notations are introduced to exclude any explicit dependence on $\xi$. One can see at low temperatures $0.03 < T^* < 0.3$ the resistivity is well described by ES law Eq.\ (\ref{eslaw}) (with $C\simeq 4.4$ at $K = 1$). The higher temperature range $1 < T^* < 200 $ is plotted separately as a function of $1/T^*$ in Fig. \ref{fig:RH}. We find two activated regimes of hopping conductivity. At high temperatures $50 < T^* < 200$, we see the large activation energy $E_a \sim E_c-\mu$, while in the range of intermediate temperatures $1 < T^* < E_g$, we see much smaller activation energy $\Delta= 0.3(E_c-\mu)$. 

The first activation energy $E_a$ does not have any physical meaning for a real SCS, because at $k_B T > E_g$ conductance of SCS is actually not due to hopping but free carriers with high energy, which are not taken into account by energy  Eq.\ (\ref{Hamiltonian}) (see Ref.~\cite{Skinner2012wit}). In contrary to $E_a$, the second activation energy  $\Delta = 0.3(E_c-\mu)$ makes full physical sense and should be seen in real experiment. The origin of this activation energy for the hopping transport is also explained in Chapter 8 of Ref.\cite{Efros1984epo}. At $T \ll E_g$, electrons optimize their conductivity by using for hopping impurities energetically close to the Fermi level. Eventually at very low temperatures, such opitmization leads to ES conductivity. However, when donor energies are slowly modulated by the long-range potential, there are large areas that do not have donors with energies close to the Fermi level and the tunneling through them is slow. Therefore, there is a range of temperatures where electrons use only nearest-neighbor donors for hopping, while activating to donors is located at the percolation level of nearest-neighbor percolation.  We then find the activation energy from the Fermi level to the nearest-neighbor percolation level by studying the hopping activation energy $\Delta$. In a heavily doped semiconductor, this energy is indistinguishable from the activation energy of electrons from the Fermi level to the conduction band percolation level $E_e$. [Of course, holes are activated from the Fermi level to their percolation $E_h$ as well so that  $\Delta =  0.3(\mu-E_h) $]. 

We verified that hopping conduction modeling correctly predicts the
activation energy of the band transport by direct calculation of the percolation level $E_e$. For this purpose, we created a cubic lattice with a small lattice constant $N_D^{-1/3}/3$. At every site of this lattice, we calculated the potential of all charged impurities and then found lowest energy $E_e$ at which percolation over this lattice takes place. The activation energy of the band transport was again close to $\Delta= 0.3(E_c-\mu)$. This result is also close to what was obtained in Ref.~\cite{Mitin2010pap}
based on an estimate of percolation level for a generic long-range random potential~\cite{Efros1984epo}. 

\begin{figure}[tb!]
\centering
\includegraphics[width=0.48 \textwidth]{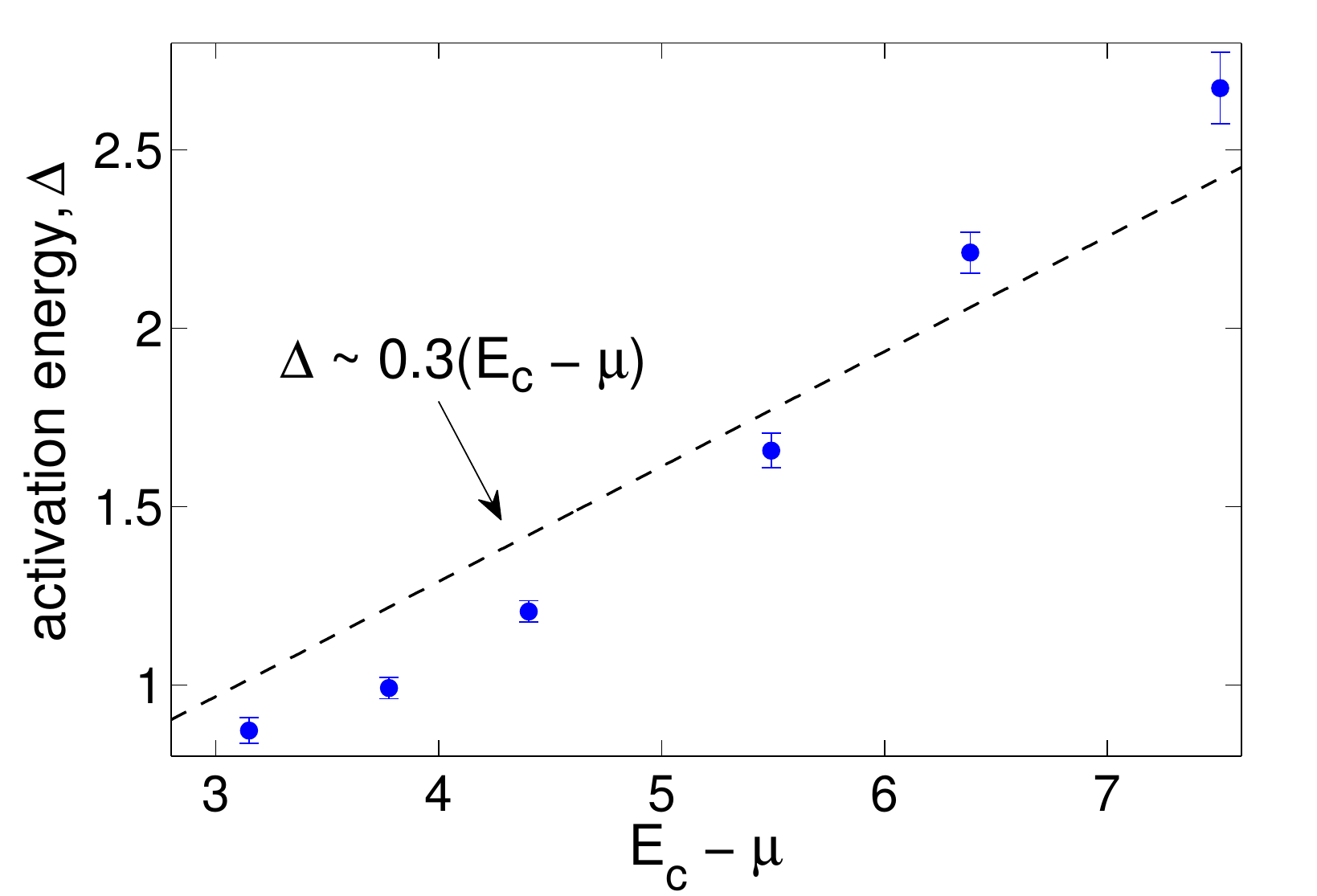}
\caption{(Color online) The activation energy $\Delta$ at $K = 1, 0.99, 0.98, 0.97, 0.96$, and $0.95$ (from right to left). The dashed line is the best linear fit $\Delta \simeq 0.3(E_c - \mu)$.}.
\label{fig:EA}
\end{figure}

In Fig.\ref{fig:EA}, we plot $\Delta$ as a function of $E_c-\mu$ for all the values $\mu(K)$ obtained at $K = 1, 0.99, 0.98, 0.97, 0.96$, and $0.95$.  
We see that the relation $\Delta \simeq 0.3(E_c-\mu)$ holds well for all $K$ in this interval.

So far, we emphasized the results that do not explicitly depend on $\xi$. Actually, a magnitude of $\xi$ is necessary to calculate $T_{ES}$. We argue now that in a TI $\xi$ is quite large leading to the prominent role of VRH. If an electron with an energy close to the Fermi level were tunneling from an electron puddle to a distant one along the straight line, it would tunnel through high barriers and its wave function would decay with $\xi \ll a_B$. Actually, a tunneling electron can use the same geometrical path as a classical percolating electron with energy $\Delta$ above the Fermi level that avoids large barriers. We assume that along such a path tunneling barriers $V$ are uniformly distributed in the range $0 \leq V \leq \Delta$ and neglect contribution of curvature of this path into action. Integration over $V$ then gives (here we return to normal units) $\xi = \hbar /(8 m \Delta/9)^{1/2}$ and $k_B T_{ES} = 4.2 (e^2/\kappa \hbar) (m \Delta)^{1/2}$.  For a TI with $a_B = N_D^{-1/3}$, we get $T_{ES} = 4.2 [(e^{2} N_D^{1/3}/\kappa) \Delta]^{1/2}$. For $\Delta$ varying between 1 and 2.5$e^{2} N_D^{1/3}/\kappa$ as shown in Fig. \ref{fig:EA}, $T_{ES}$ changes from 4.2 to 6.6$e^{2} N_D^{1/3}/\kappa$. For $\kappa = 30$, $N_D = 10^{19} cm^{-3}$, and $e^{2} N_D^{1/3}/\kappa k_B \simeq$ 100\ K, $T_{ES}$ varies from 420 to 660\ K. In order to study VRH in TI samples experimentally, one has to deal with large enough samples, where surface conductance is smaller than the bulk one.
\footnote{Historically VRH between puddles was studied in Ref.~\cite{Shklovskii1973Sov}. This paper was written before Ref.~\cite{Efros1975cga} and claimed Mott VRH. Now it is clear that resistivity obeys Eq.\ (\ref{eslaw}). The theory~\cite{Shklovskii1973Sov} of the transition from activated transport to ES law is to be modified as well, but we are not dwelling on this transition range, because it is difficult to study details of such a transition in experiment.}

\section{Thermopower}

In the recent paper~\cite{Akrap2012tp}, the authors studied activation energy of the bulk resistivity of series of samples of Bi$_{2}$Te$_{3-x}$Se$_{x}$ with different $x$ and thereby different positions of the Fermi level in the TI gap. 
They found that when the Fermi level sinks into the gap, the activation energy of resistivity $\Delta$ grows and reaches a maximum at 40 meV and then decreases. The increase of the activation energy $\Delta$ on both sides of the maximum is accompanied by the increase of the absolute value of the thermopower $S$. However, near the maximum of $\Delta$, the thermopower abruptly changes its sign. These findings are in agreement with what one can expect when a semiconductor goes through the point of complete compensation. Here, we would like to concentrate on the maximum absolute value of the thermopower, for example, at $n$-type side of the maximum. 

It is known that for flat bands $n$-type semiconductor with the Fermi level $\mu$ inside its gap the thermopower $S = \Delta/eT$ , where the activation energy $\Delta = E_c -\mu$. For bended bands of a strongly compensated $n$-type semiconductor, one could think that $S = \Delta/eT$, where the activation energy $\Delta = E_e -\mu$ is determined by the activation to percolation level $E_e$. Actually, it was argued~\cite{Overhof1981pm, Quicker1999iod,Overhof2000eol} that the Peltier energy (heat) $\Pi = eTS$ is determined by the average potential energy of electrons $E$ (conduction band bottom) along most conducting one-dimensional percolation paths,  $\Pi = <E-\mu>$. (We call a percolation path any line where the potential energy of electron is smaller than $E_e$ and we call a set of the least resistive of these paths, which carry most of the current, the most conducting percolation paths.) The thermopower of an open circuit following an individual percolation path can be obtained by integrating $E-\mu$ along this path. Among two parallel paths connecting points A and B, the more resistive one has a somewhat larger open circuit thermopower and, therefore, drives circular current back through the least resistive one. This current reduces thermopower of the resistive path so that the voltage between A and B is determined by the more conducting path.

If the probability distribution of potential energy $E$ on most conducting paths is the same as for the unconditional probability distribution of $E$, which we call DOS $g^*(E)$ above, we can use $g^*(E)$ to calculate  $\Pi$ and $S$. For example, in the case of a constant  $g^*(E)$ for $\mu < E < E_e$, we get $E_S = <E-\mu> = \Delta/2 = (E_e -\mu)/2$. This conclusion was confirmed by the numerical experiment~\cite{Overhof1981pm} for the case of a constant $g^*(E)$.

In a strongly compensated semiconductor, one can use the real $g^*(E)$ found above. For example, at $K=0.95$ one can use Fig. \ref{fig:EA} to find that $\Delta = E_e - \mu \simeq 1$. Then using DOS shown in Fig. \ref{fig:DOS} one can check that the average energy in the range of $0 < E < 1$ is $<E - \mu> \simeq \Delta/2 = 0.5$. Thus our simple approximate prediction is that the largest achievable $\Pi \simeq \Delta/2$. This conclusion is valid for all $K \leq 0.98$ we studied. 

For the data of the paper~\cite{Akrap2012tp}, our prediction means that at $T=100$ K the largest thermopower $S =\Pi/eT$ observed should be of the order 25 mV/100 K = 0.25 mV/K in resonable agreement with the observed value $S=0.4$ mV/K.

In this paper, we are not considering the additional contribution to thermopower of activated electrons from phonon drag~\cite{Gurevich,Herring1954tot}. This effect becomes significant only at temperature $T \leq T_D/3$, where $T_D$ is the Debye temperature, because at larger temperatures, the low-energy phonons interacting with electrons are strongly scattered by thermal phonons,
which in turn are strongly interacting with imperfections of the crystal. In Bi$_{2}$Se$_{3}$, $T_D \sim 150$ K, so that phonon drag should get important only below 50 K (where electron transport is already via hopping), while the activated transport we are interested in happens at $T \geq 100$ K.

In order to go beyond the above approximation that the distribution of energies on paths contributing to $\Pi$ is given by the density of states $g(E)$, we calculate currents $I_{ij}$ in every Miller-Abrahams resistor $R_{ij}$ and the total current $I(U)$ for a small applied voltage $U$ by solving Kirchhoff equations for the ground state of impurities obtained by our algorithm. Following Ref.~\cite{Zvyagin1973ott}, we then calculate the energy flux through a cross-section of the sample $Q(U)$ as a sum of energy fluxes carried by resistors $q_{ij} = (E_i + E_j)I_{ij}/2e$ and found $\Pi = Qe/I$. We simplify the implementation of this procedure by modifying our algorithm in the following way: instead of dealing with completely randomly positioned donors and acceptors, we randomly position them on all sites that are appropriate to their number cubic lattice. To find the energies $E_i$, we use a simple Coulomb potential. (There is no need in truncation at small distances via finite $a_B$.) We concentrate on the range of relatively high temperatures, where the conductivity is characterized by activated behavior. We checked that the conductance $I/U$ has the same activation energy $\Delta$ as obtained by the percolation algorithm. We found that in the range of $0.95 \leq K \leq 0.98$, where the asymmetry of the density of states is large and donors dominate the transport, Peltier energy $\Pi/\Delta \simeq 0.40 \pm 0.05$, not too far from the simplified theories and the experimental data~\cite{Akrap2012tp}. For $K > 0.98$, growing donor-acceptor symmetry reduces $\Pi$ and brings it to zero at $K=1$, in agreement with the data of the paper~\cite{Akrap2012tp}.

\section{Conclusion}

In this paper, we applied the model of strongly compensated semiconductor to a bulk TI with narrow gap. We calculated the activation energy of the bulk resistivity $\Delta$ and showed that it grows as $\Delta = 0.3 (E_c-\mu)$, when the compensation degree $K \rightarrow 1$ and the Fermi level sinks into the gap. If one of the two carriers still dominates and the thermopwer is still monopolar the Peltier energy is $\Pi \simeq \Delta/2$. Both predictions seem to agree with most of the TI data. 

We would like to mention that the same model is able to interpret measurements of the Hall Effect obtained for the same samples. The  Hall constant $R_H$ is expected to grow exponentially with decreasing temperature with the same activation energy $\Delta$ as the resistivity~\cite{Karpov1982Sov,Shik1995epo,Overhof1981pm}. The reason for such growth is that $R_H$ is dominated by nodes of percolation path network that occur at energy close to the percolation level. Such nodes are relatively rare at low temperatures. Therefore $R_{H}(T) = \rho(T)u(T)/c$ grows with decreasing $T$, where mobility $u(T)\propto T^{m}$ and $m \geq 2$. The observed behavior of $R_{H}(T)$ does not contradict this prediction~\cite{Ren2011o$s}. Indeed, the largest activation energy of $R_H$ was found to be on average $\sim 15$ meV larger than the largest $\Delta \sim 50$ meV. This difference is of the order of $1.5k_B T$ at the characteristic measurement temperature of activation law $T = 100 K$ and, therefore, the experimental data is compatible with a power law $u(T)$. In future work, we plan to narrow the range of theoretical predictions by a numerical evaluation of $R_H$ for the simulated above potential of our model. 
     
\begin{acknowledgments}

We are grateful to Brian Skinner with whom this work was started,
for reading the manuscript and helpful advices.
We acknowledge useful discussions with A.\ Akrap, Y.\ Ando, A.\ L.\ Efros, Y.\ M.\ Gal'perin, J.\ Kakalios and I.\ P.\ Zvyagin. T. Chen was partially supported by the FTPI.

\end{acknowledgments}

\bibliography{Thermopower}

\end{document}